\documentclass[doublecol, figures,maxnames=1]{epl2}
\usepackage{amsmath}
\usepackage{hyperref}
\usepackage{ulem}
\begin{document}

\title{
Magnetic vortices induced by a moving tip
}
\author{Martin P. Magiera \inst{1}\thanks{E-mail:
    \email{martin.magiera@uni-due.de}} \and Alfred Hucht \inst{1} \and
  Haye Hinrichsen \inst{2} \and Silvio R. Dahmen \inst{3} \and
  Dietrich E. Wolf \inst{1}}

\shortauthor{M. P. Magiera \etal}
\institute{
\inst{1} Faculty of Physics and CeNIDE, University of Duisburg-Essen,
D-47048 Duisburg, Germany, EU\\
\inst{3} Universit\"at W\"urzburg, Fakult\"at f\"ur Physik und Astronomie,
97074 W\"urzburg, Germany, EU\\
\inst{2} Instituto de Fisica, Universidade Federal do Rio Grande do Sul,
91501-970 Porto Alegre RS, Brazil
}
\date{\today}

\abstract{A two-dimensional easy-plane ferromagnetic substrate,
  interacting with a dipolar tip which is magnetised perpendicular
  with respect to the easy plane is studied numerically by solving the
  Landau-Lifshitz Gilbert equation. Due to the symmetry of the dipolar
  field of the tip, in addition to the collinear structure a magnetic
  vortex structure becomes stable. It is robust against excitations
  caused by the motion of the tip.  We show that for high excitations
  the system may perform a transition between the two states. The
  influence of domain walls, which may also induce this transition, is
  examined.}
\maketitle
\section{Introduction}
Vortices in magnetic layers have been intensely studied
\cite{Belavin1975, Nikiforov83,
  Huber82,Gouvea1989,Shinjo2000,Wachowiak2002,Costa2011}. They are
topological defects, and can be characterised by a chirality (an
integer winding number, which is a conserved quantity for topological
reasons) and a polarity (the out-of plane magnetisation).  They can
annihilate by antivortices having a winding number of opposite
sign. Vortex-antivortex pairs are important excitations in two
dimensional magnetic systems. In open systems one can also excite
isolated vortices, when the corresponding antivortex leaves the system
at the boundary.

The polarity of a vortex may be treated as a bit, as it possesses two
very stable states which can be easily probed \textit{e.g.}\ with GMR
sensors, as those used in reading heads of magnetic disks. Thus
magnetic vortex structures are promising candidates for novel
non-volatile storage concepts.  An important question in this context
is, how to ``write'' a vortex.  To switch the polarization, magnetic
field pulses \cite{Xiao2006, Hertel2007}, alternating magnetic fields
\cite{Waeyenberge2006} or spin-polarised currents \cite{Yamada2007}
have been proposed.

In this letter we present and analyse a new method, by which vortices
can be generated or removed.  The excitation energy is provided by a
moving magnetic tip as used in magnetic force microscopy 
  (MFM).  It is known that MFM tips do influence the substrate, which
  they are supposed to probe -- an undesired effect for the microscopy
  purpose. It has been observed in experiments that domain walls are
  deformed by the passing tip \cite{Mamin1989,Lee2011}. Tip controlled
  domain wall manipulation has been achieved
  \cite{Imre2003,Yamaoka2006}. The manipulation of vortices by an MFM
  tip has been realized in a type-II superconductor
  \cite{Auslaender2008,Brandt2010}. Thus, it is of great interest from
  the microscopy and the manipulatory point of view to study how an
  MFM tip interacts with the scanned surface.

We first show that vortex states are stable configurations in easy
plane ferromagnetic structures in the presence of a magnetic tip,
positioned above the ferromagnetic structure. Such a system has been
studied recently in order to explore the friction force decelerating
the magnetic tip \cite{Magiera09a, Magiera11, Magiera11b}. Then we
show that the vortex remains stable when the tip is moved along the
substrate, dragging the vortex through the substrate. Moreover the
moving tip may also create or destroy a vortex structure, depending on
the tip magnetisation and velocity, offering an alternative way to
switch between 
three states: A collinear state, as well as a vortex state with up or
down polarity. Finally we study the stability of the dragged vortex
structure when it passes through a domain wall.

\section{The system}
Our system consists of $N{=}L_x{\times}L_y$ classical Heisenberg
spins $\mathbf S_i {=} \boldsymbol{\mu}_i/\mu_s$ on a square grid, where
$\mu_s$ is a material specific saturation magnetisation. The
Hamiltonian contains two terms, corresponding to a substrate and a tip part
\begin{equation}
\mathcal H = \mathcal H_\mathrm{sub} + \mathcal H_\mathrm{tip},
\end{equation}
For the spin-spin interaction in the substrate we assume an isotropic
exchange with interaction constant $J$ and equivalent easy axes in
$x$- and $y$-direction,
\begin{equation}
\mathcal H_\mathrm{sub} = -J \sum_{\left < i,j \right >} \mathbf S_i
\cdot \mathbf S_j - d_z \sum_i S_{i,z}^2 - e_{xy} \sum_i
(S_{i,x}^4 + S_{i,y}^4).
\label{eq2}
\end{equation}
The easy-plane anisotropy $d_z{<}0$ alone would lead to an infinite domain wall
width, as all in-plane magnetisation configurations would be
degenerated. Correspondingly, the Mermin-Wagner theorem would rule out long
range magnetic order at finite temperatures \cite{Mermin1966}.
The fourth-order anisotropy term $e_{xy}{>}0$, however, breaks the
continuous symmetry, so that the Hamiltonian \eqref{eq2} has a
ferromagnetic low temperature phase with domain walls of finite width. 
We use the anisotropy parameters $d_z {=} {-}0.1J$ and
$e_{xy}{=}0.1J$ in this letter.

\begin{figure}[t]
\centering
\includegraphics[width=.49\columnwidth]{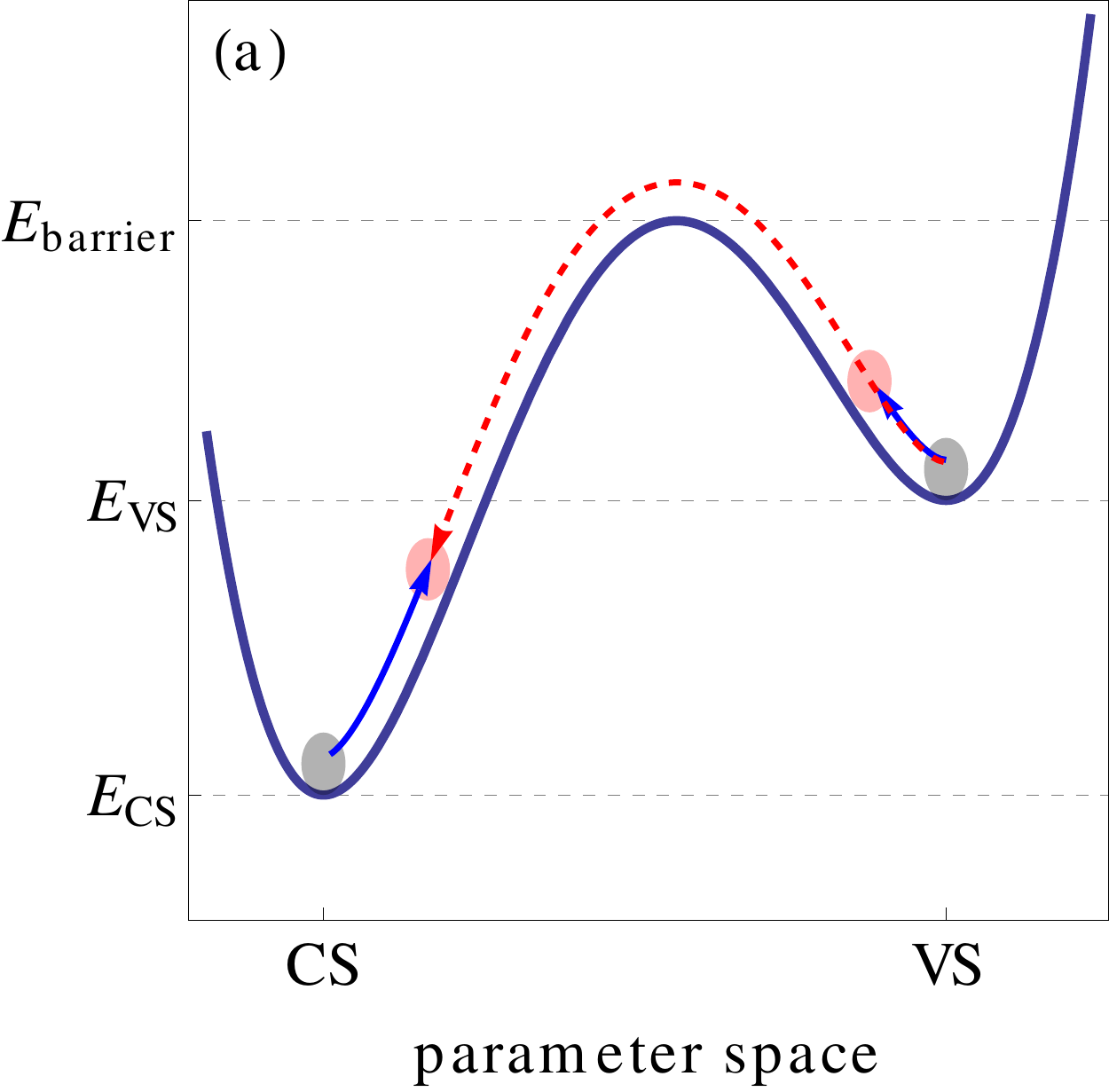}
\includegraphics[width=.49\columnwidth]{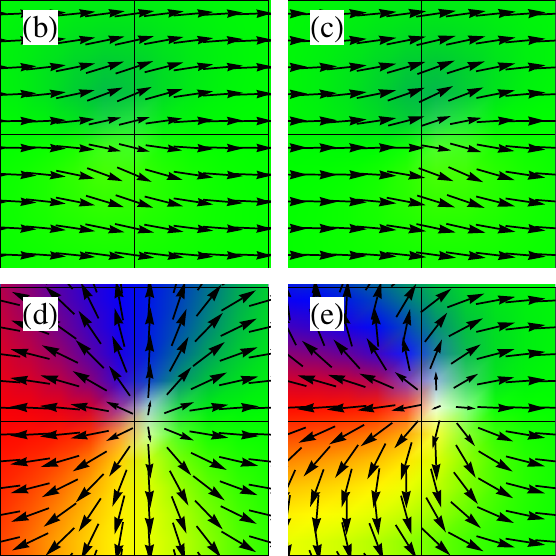}
\caption{\label{fig:System_near_eq}(Colour on-line) (a) Sketch of the
  energy landscape. At least two local minima, corresponding to the CS
  (collinear state) and the VS (vortex state), exist. The absolute
  minimum depends on the tip strength $w$. There is a potential
  barrier between the two states and thus the CS and the VS may be
  very stable against external driving (blue arrows). A strong
  perturbation may lead to the transition from the VS to the CS (red
  dashed arrow) and vice versa.  (b-e) A part of a system near
  equilibrium, initialised in the CS (b,c) or the VS (d,e). The
  scanning velocities are $v=0.01$ (b,d) and $v=0.3$
  (c,e). The colour coding as well as the arrows represent the
  magnetisation in the $xy$-plane, \textit{cf.}\
  fig.~\ref{fig:DW_eq}.}
\end{figure}
The substrate-tip interaction is introduced via a dipolar term,
\begin{equation}
 \mathcal H_\mathrm{tip} = - w \sum_i {\frac{3
  ~(\mathbf S_i \cdot \mathbf e_i) (\mathbf S_\mathrm{tip} \cdot
  \mathbf e_i) - \mathbf S_i \cdot \mathbf S_\mathrm{tip} } 
  {R_i^3}}, 
\label{eq:htip}
\end{equation}
where $R_i = \left |\mathbf R_i \right |$ denotes the norm of the
position of spin $i$ relative to the tip $\mathbf R_i = \mathbf r_i -
\mathbf r_\mathrm{tip}$, 
and $\mathbf e_i$ its unit vector $\mathbf e_i = \mathbf R_i / R_i
$. $\mathbf r_i$ and $\mathbf r_\mathrm{tip}$ are the position vectors
of the substrate spins and the tip respectively.  $w$ is a free parameter that
quantifies the
dipole-dipole-coupling between the substrate spins and the tip, thus controlling
the strength of the tip. We use
$\mathbf{S}_\mathrm{tip}=(0,0,-1)$, $w$ is a free parameter representing the
tip strength. The tip is moved with constant velocity $(v,0,0)$ two
lattice constants above the substrate, along the middle line between
two spin rows. 

For a real system the tip is of course not a point dipole
  as represented by \eqref{eq:htip}. It rather resembles a magnetic
  cone of micron length. Depending on the scanned length scales the
  distance between surface and probe, as well as on the length and
  shape of the probe, different approximations of the tip field are
  used, see Ref.~\cite{Haberle2012} and references therein. While the
  dipole approximation yields the correct far-field behavior, it has
  been shown in Refs.~\cite{Thiaville1997,Hug1998,Garcia2001} that
  the stray field of a hollow cone-type tip may be approximated by a
  magnetic point charge, if the tip extension is large compared to the
  distance between tip and surface. As the tip magnetization is
  assumed perpendicular to the surface in this work, both the dipole
  and the monopole approximation favor a vortex directly underneath
  the tip. The effects described in the following are therefore
  expected to be qualitatively similar in both cases, however they may
  be more pronounced for the monopole field due to its longer range.

Two different kinds of boundary conditions will be used. First we
consider the case that all inhomogeneities in the system are due to
the tip. Then it is natural to describe the system in the comoving
frame of the tip \cite{Magiera09a}: In the $y-$direction, the
boundaries are open, in the $x-$direction dynamical. When the tip
advances by exactly one lattice constant, the foremost row is
duplicated, and the last one is deleted. That way, arbitrarily long
times can be simulated. In the second part we investigate, how vortex
generation is influenced by a domain wall that is pinned far away from
the tip. In this case, a combination of open ($x-$direction) and fixed
($y-$direction) boundary conditions is more appropriate.  The system
sizes used are $64\times 48$ for the case of simulations in the
absence of domain walls, and $200 \times 48$ for the case where domain
walls are of interest.

The equation of motion is the Landau-Lifshitz-Gilbert (LLG) equation
\cite{LandauLifshitz1935, Gilbert2004}
\begin{equation}
 \frac{\partial}{\partial t} \mathbf S_i = -\frac{\gamma}{(1 +
  \alpha^2) \mu_s} \left[\mathbf S_i \times \mathbf h_i + \alpha ~
  \mathbf S_i \times(\mathbf S_i \times \mathbf h_i) \right],
\end{equation}
with saturation magnetisation $\mu_s$, gyromagnetic ratio $\gamma$,
the phenomenological damping constant $\alpha$ (we use the high
damping value $\alpha{=}0.5$ in this letter to reach a steady state in
a short simulation time) and the local field $ \mathbf h_i =
- \partial \mathcal H/\partial \mathbf S_i.  $ It produces Larmor
precession with frequency $\left | \mathbf h_i\right| \gamma/\mu_s$,
and a damping in the direction of the local field. In equilibrium,
each spin points in the direction of its local field. To solve the LLG
we use the Heun integration scheme \cite{Garcia98}.

\section{Nonequilibrium steady states}
\begin{figure}[bt]
\centering
\includegraphics[width=.7\columnwidth]{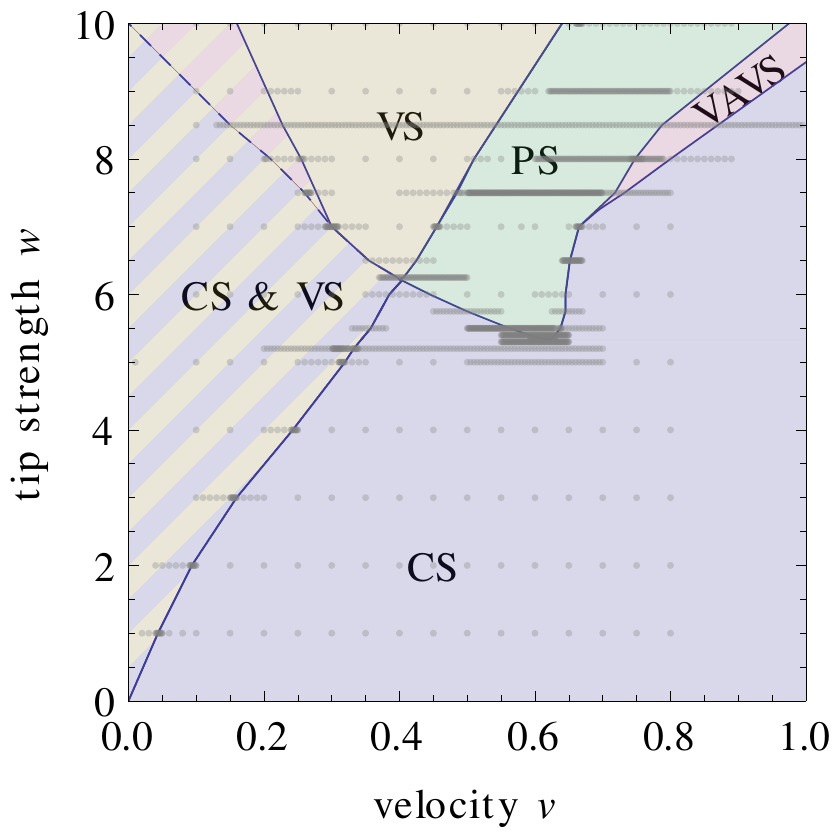}
\caption{\label{fig:NESS_DIAG}(Colour on-line) Nonequilibrium steady state diagram with
  the pure states (collinear state, CS, and vortex state, VS), the
  co-existence state (CS \& VS), the periodic state (PS) and the
  vortex-antivortex state (VAVS). The grey points represent the underlying
  simulations.}
\end{figure}
Before studying the driven system, let us discuss the equilibrium case
($v{=}0$). If the tip is absent ($w{=}0$), the equilibrium
configuration corresponds to all spins pointing in the same direction,
the collinear state (CS). 

As we increase the tip strength, a second minimum appears in the
potential landscape, the cylindrically symmetric vortex state (VS,
\textit{cf.}\ fig.~\ref{fig:System_near_eq} (d)). At the same time,
the CS minimum is continuously moved to the configuration of a
slightly disturbed collinear configuration (\textit{cf.}\
fig. \ref{fig:System_near_eq} (b)). For any reasonably small tip
strength, both equilibrium states, the CS and the VS, are stable. If
we initialise a VS, and switch off the tip, the VS is still stable:
Being a topological defect, its winding number is a conserved quantity
which cannot spontaneously change.

\begin{figure}[tb]
\centering
\includegraphics[width=\columnwidth]{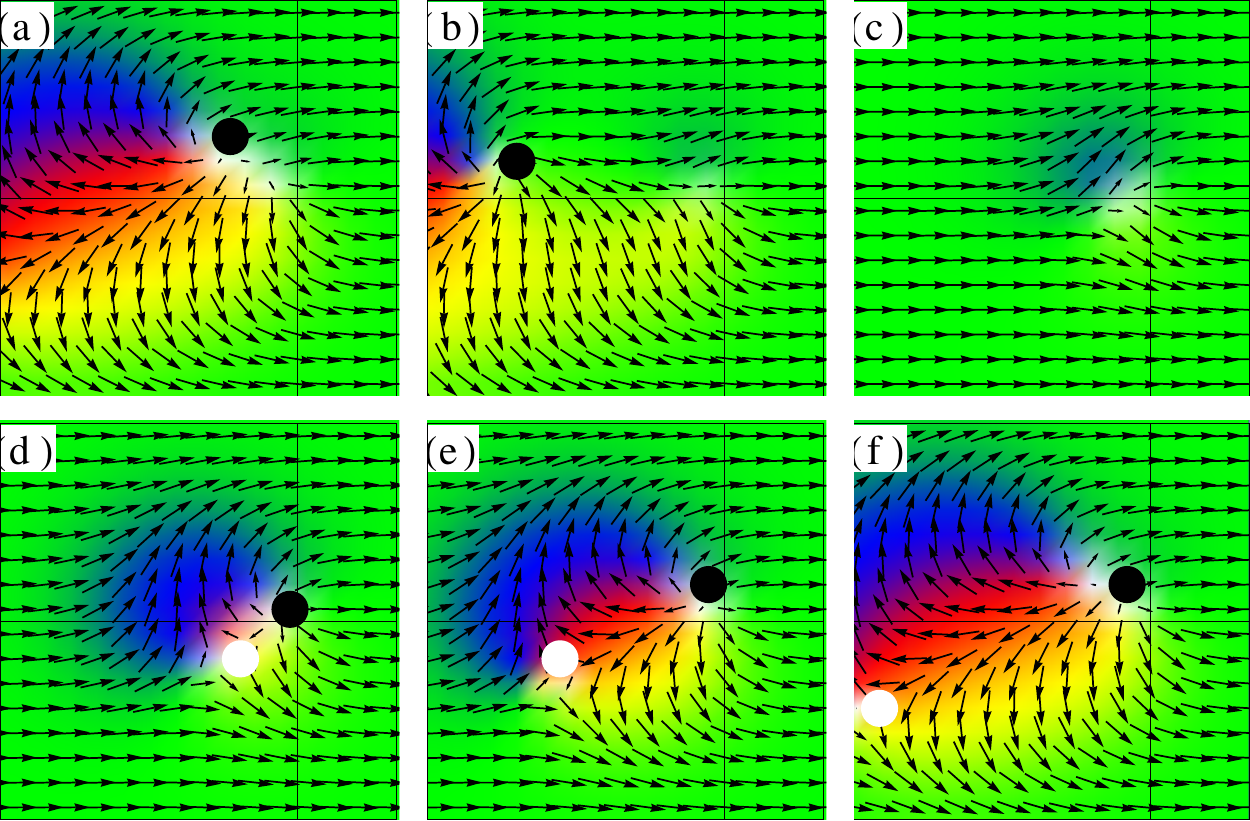}
\caption{\label{fig:PS_TS}(As a movie and colour on-line) Some
  snapshots of a system in the PS in a co-moving frame of reference,
  the tip positioned at the origin (the two lines correspond to the
  $x$- and $y$-axis). Vortex cores are marked as black dots,
  antivortex cores as white dots. Here we observe the periodical
  transition from the VS to the CS (a-c) via the release of the vortex
  from the tip, as well as the transition from the CS to the VS (c-f)
  via a VAVP creation. See the complete cycle as a movie online.}
\end{figure}
Let us now move the tip with constant velocity ($v{>}0$): The system
is driven into a nonequilibrium steady state (NESS), which is near the
equilibrium for small excitations. Accordingly, in the NESS diagram
(\textit{cf.}~fig.~\ref{fig:NESS_DIAG}) we see at $v{>}0$ a region
where the CS and the VS coexist. Here the initial configuration
determines the finally stabilised NESS. As the excitation overcomes a
threshold by increasing the tip strength $w$ or scanning velocity
$v$\footnote{A value $w=1$ in our model corresponds to a magnetic
  field of about $0.1$T in the substrate under the tip.}, the system
may perform a transition from the VS to the CS (when the CS represents
the total minimum) or vice versa, as sketched in
fig.~\ref{fig:System_near_eq}(a) by the red dashed arrow. In the NESS,
we then observe either the pure CS or the pure VS.

Another mixed state is the periodic state: Here the system is excited
so strongly that neither the VS nor the CS is stable. The system then
flips between the states back and forth continuously. We explain this
state in more detail, because it sheds also more light onto the
transitions to the pure states mentioned above. Let us start in the
VS. Here the vortex is bound by the tip, because the cylindrically
symmetric structure minimises the tip energy. However, in the PS (or
when the CS is the corresponding NESS) the tip pumps so much energy
into the substrate that the vortex may decouple from the tip, and
moves away towards the system boundary (fig.~\ref{fig:PS_TS}(b)),
leaving a CS behind (fig. \ref{fig:PS_TS}(c). Accordingly the potential
barrier in fig.~\ref{fig:System_near_eq}(a) corresponds to the energy
of a free moving vortex and a CS. If the system's NESS is the CS, the
system has now reached its steady state. If the initial state was the
CS, but the corresponding NESS is the VS or the PS, the system is
again excited further by the tip, and a vortex-antivortex pair (VAVP)
may nucleate under the tip (fig.~\ref{fig:PS_TS}(d)). The energy of a
VAVP represents the potential barrier which the system has to
overcome at this transition. After the nucleation, the antivortex
moves away from the tip. It may stay at a constant distance (\textit{e.g.}\
fig.~\ref{fig:PS_TS}(e)), and we end up in a NESS called
vortex-antivortex state (VAVS). If the antivortex moves out of the
system, we again get a VS (fig.~\ref{fig:PS_TS}(f)).

The two transitions lead to very different paths in configuration
space. The reason why two different barriers have to be passed, as
well as why one barrier is higher than the other, can be found in the
topology of the system: For the transition CS$\rightarrow$VS, a vortex
must be created. However, an isolated vortex cannot be created, as it
represents a topological defect, which violates vorticity conservation
(the total number of vortex and antivortex cores). Only VAVPs can be
created, and thus the energy barrier here is higher than that of the
VS$\rightarrow$CS transition. The reason why the latter transition
(which violates vorticity conservation) can occur at all, is that here the
antivortex core interacts with the open system boundary. The
different transition paths of the PS lead to a hysteresis.

\section{Influence of domain walls}
\begin{figure}[bt]
\centering
\includegraphics[width=.38\columnwidth]{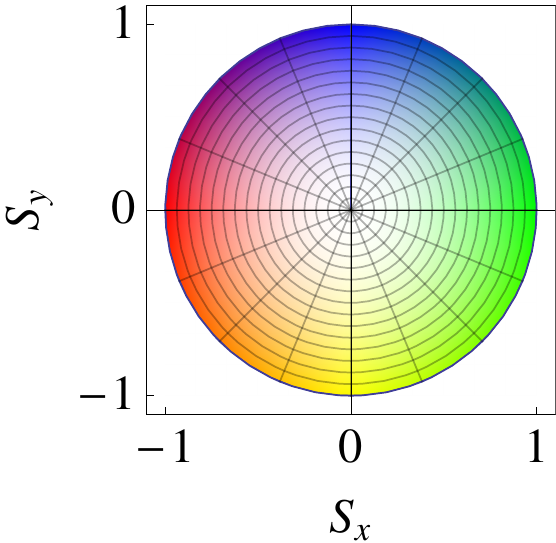}
\includegraphics[width=.6\columnwidth]{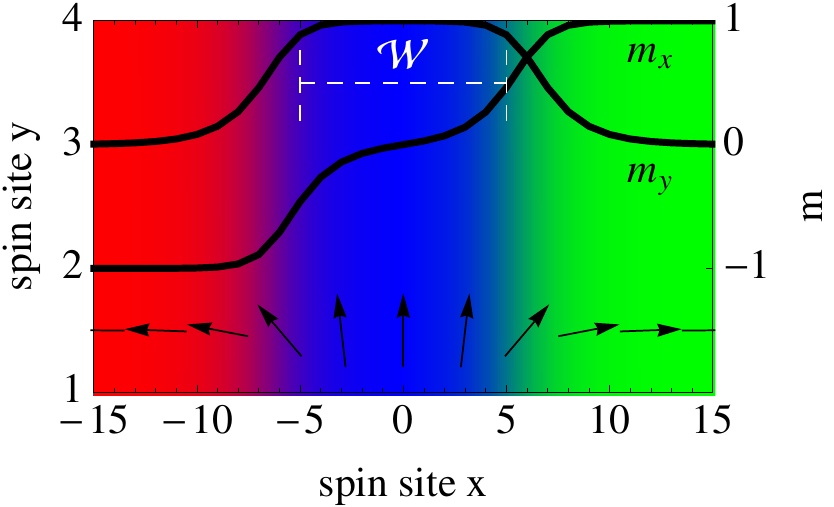}
\caption{\label{fig:DW_eq}(Colour on-line) Legend for the colour coded plots, and the
  equilibrium configuration of a system with two domain walls, as it
  is used for the simulations as an initial state. The width of the
  middle domain $\mathcal W$ is a free parameter.}
\end{figure}
In the section above we studied a perfect system in order to get an insight into the
occurring NESSs.  However, a real magnetic specimen contains domain
walls, which the tip has to pass through. The open 
question is, how stable the above characterised NESSs are against
\textit{e.g.}\ domain walls. In order to answer this question we now initialise a system which
contains two $\pi/2$ domain walls in equilibrium, where the distance
between the two domain walls (or the length of the middle domain) is a
free parameter $\mathcal W$ (\textit{cf.}\ fig.~\ref{fig:DW_eq}). Additionally
we fix the boundaries in $y$-direction by the equilibrium
magnetisation, as it is sketched in fig.~\ref{fig:DW_eq}, to emulate
infinitely long domain walls.
\begin{figure*}[t]
\includegraphics[width=\textwidth]{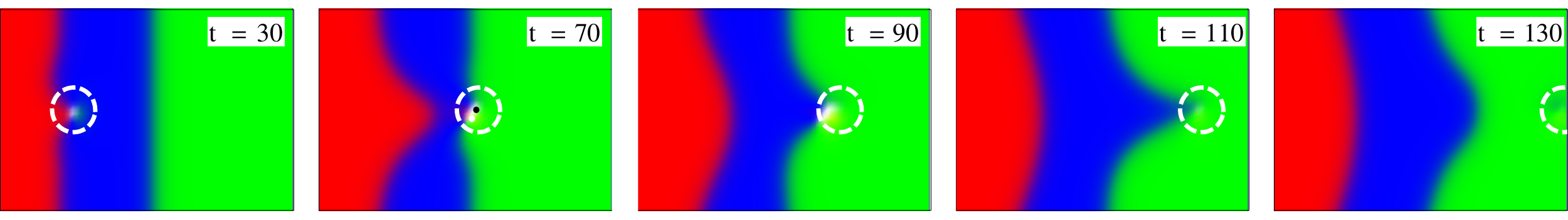}
\caption{\label{fig:weakexcitation}(As a movie and colour online) Fast
  magnetic tip (marked by the dashed circle, $w=3, v = 0.5$),
  interacting with two domain walls, with the domain wall distance
  $\mathcal W = 20$. The snapshots show a part of the whole system
  containing $50^2$ spins. The high excitation may lead to the
  creation of a VAVP, see (c). As the NESS of the system is a CS, the
  VAVP is annihilated again, leaving the initial domain wall
  configuration.}

\end{figure*}
\begin{figure*}[bth]
\centering
\includegraphics[width=\textwidth]{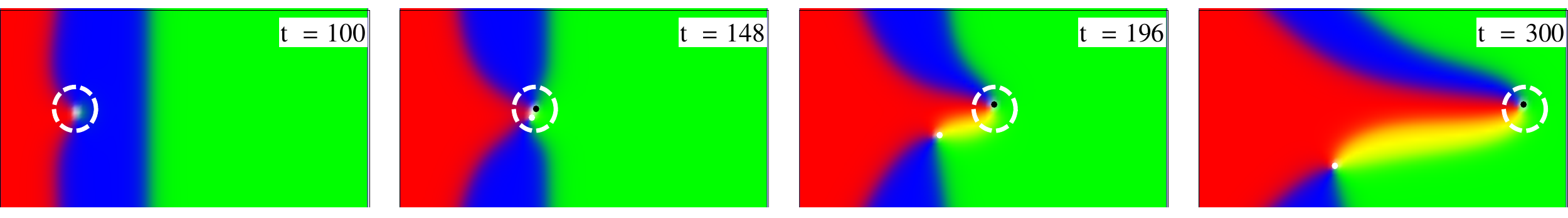}
\caption{\label{fig:strongexcitation}(As a movie and colour online) Slow, but
  strong magnetic tip ($w=5, v=0.3$), interacting with two domain
  walls. Again only a part of the system ($70\times50$ spins) is
  plotted. A VAVP is created. As the NESS for the present $v-w$
  combination is the VS, the vortex stays bound by the tip, and the
  antivortex travels out of the system, leaving a VS instead of the
  domain wall state.}
\end{figure*}
The initial condition corresponds to the simulations initialised in
the CS above, as we always start in the left domain. Let us start with
a weak excitation ($w {\rightarrow} 0$). Then the domain wall
configuration is not disturbed by the scanning tip. We may observe a
slight bending when the tip just passes by (like in the very left panel of
fig.~\ref{fig:weakexcitation}), which soon relaxes again,
like a rubber band.  After the tip has passed the domain walls the
configuration is again in its initial state.

When the perturbation by the tip is stronger, a VAVP may nucleate at
the domain walls, even if the energy of the moving tip alone is not
sufficient, as the domain walls provide additional energy
(\textit{cf.}\ fig.~\ref{fig:weakexcitation}).  Which state is finally
adopted depends on the NESS. If the corresponding state is the CS, the
VAVP get annihilated, and the initial domain wall state appears
again. If the corresponding state is the VS, the distance between the
vortex and the antivortex increases. Finally, the antivortex stays at
the system boundary (as we have fixed boundary conditions in this part
the antivortex cannot leave the system), leaving a VS that trails
along with the tip.  We observed that in the coexistence regime (CS \&
VS) the system ends up in the VS.  The PS is not influenced by the
domain wall. It may occur that at the domain walls additional VAVPs
nucleate, which after a short lifetime annihilate again.

The influence of the domain wall distance $\mathcal W$ is the
following: The smaller $\mathcal W$, the larger is the energy density,
the domain wall may provide to a VAVP creation process. Accordingly,
the lowest tip strength $w$, at which a VAVP is created, is larger for
larger $\mathcal W$.

Finally we discuss the influence of the boundary conditions. For
instance from the right panel of fig.~\ref{fig:strongexcitation} one
may claim that the fixed boundaries have a strong impact on the
stability of the VS, and enforce the annihilation of the antivortex,
as the separation of the VAVP generates continuously growing domain
walls. These may become arbitrarily large and thus energetically much
more unfavorable than the domain wall state. To get an annihilation
process, vortex and antivortex must first move toward each other.
This is only possible when the vortex is released from the tip first,
as a free moving vortex or antivortex cannot move at velocities
comparable to that of the tip. If the vortex is released or not from
the tip depends solely on the current NESS, and not \textit{e.g.}\ on
the tailing domain walls, as they cannot provide additional energy
density under the tip for a vortex release event. All observed VAVP
annihilation processes occur directly after the tip-domain wall
interaction, where the events under the tip may be seen as completely
decoupled from the walls.

\begin{figure}
\includegraphics[width=.49\columnwidth]{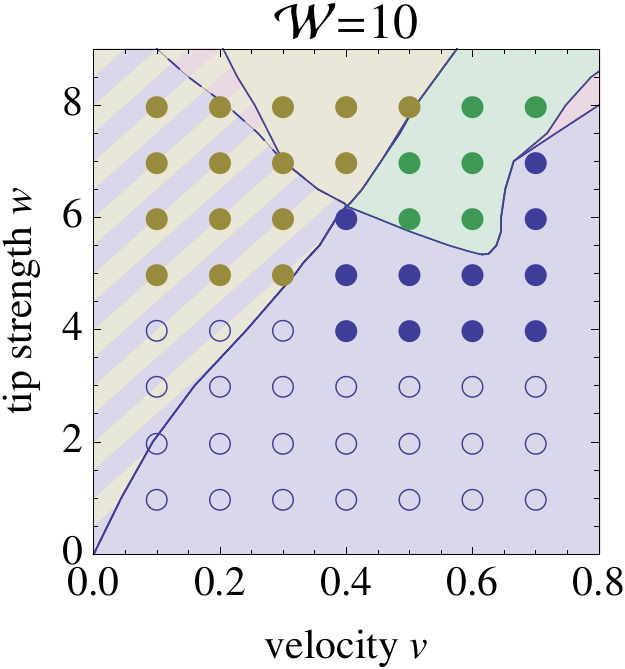}
\includegraphics[width=.49\columnwidth]{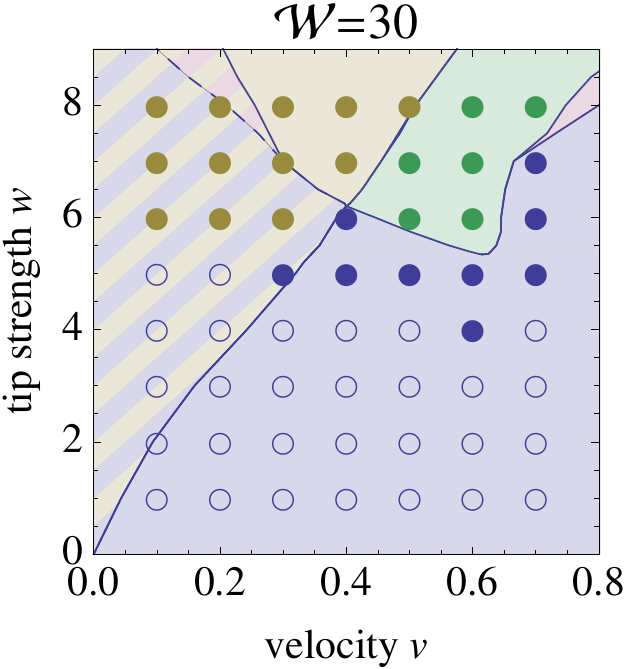}
\caption{(Colour on-line) Influence of domain walls on the NESS. Empty circles
  represent simulations, where the system does not create any VAVP,
  filled circles represent these simulations where at least one VAVP is
  created. The colour coding denotes which final state has been
  observed: Blue circles represent the CS, yellow ones the VS, and
  green ones the PS.}
\end{figure}

\section{Conclusion}
A tip, which is aligned above a ferromagnet with an easy plane
anisotropy, and magnetised perpendicular to the easy plane,
energetically stabilises a vortex state in the substrate. The vortex
state is stable against slight perturbations, occurring \textit{e.g.}\
when the tip is moved with constant velocity $v$ parallel to the
substrate. At a threshold velocity, which depends on the tip
magnetisation, the substrate may perform a transition from the vortex
phase to the collinear phase and vice versa, but it may also exhibit a
periodic switching for strong excitations.  In a real
  system, $w$ is replaced by an effective value, which varies with the
  characteristics of the tip and the distance between the tip and the
  substrate, because both parameters determine the field acting at the
  surface and thus the energy injected into the system.  In summary
we have three possible states the system may adopt (the CS state and
the VS state with up or down polarity), and which may be switched by a
moving tip.  The presence of a domain wall effectively shifts the
state separation line between the coexistence state and the pure
vortex state to lower tip magnetisation values.

\begin{acknowledgments}
  We thank Sebastian Angst for valuable
  discussions.  This work was supported by the German Research
  Foundation (DFG) through SFB 616 ``Energy Dissipation at
  Surfaces''. A part of the work has been performed during a stay in
  Porto Alegre, Brazil, granted by the German Exchange Association
  (DAAD) through the Project Related Exchange Brazil-Germany
  (PROBRAL).  Computing time by the Neumann Institute for Computing
  (NIC) is gratefully acknowledged.
\end{acknowledgments}

\bibliographystyle{eplbib}
\bibliography{paper}

\end{document}